\documentstyle[12pt,epsf]{ioplppt}  
\newcommand{\postscript}[2] {\setlength{\epsfxsize}{#2\hsize}
\centerline{\epsfbox{#1}}}
\newcommand{\be}{\begin{equation}}
\newcommand{\ee}{\end{equation}}

\begin{document}

\jl{4}

\title{Lower bound to the mass of a relativistic three-boson system
in the light-cone}
%[Bound states in the light-cone]

\author{J.P.B.C. de Melo$^{\dag,*}$, A.E.A. Amorim$^\S$,
Lauro Tomio$^{\dag}$ \ and T. Frederico$^\ddag$} 

\address{$^\dag$ 
Instituto de F$\acute{\imath}$sica Te\'orica, Universidade Estadual
Paulista, 01405-900, S$\tilde{a}$o Paulo, SP, Brazil}
\address{$^*$ 
LPNHE/LPTE Universit\'e Pi\'erre \& Marie Curie,   
4 Place Jussieu, F-75252 Paris Cedex 05, France}
\address{$^\S$ 
Faculdade de Tecnologia de Jahu, CEETEPS, Jahu, Brazil} 
\address{$^\ddag$ Departamento de F$\acute{\imath}$sica, ITA, Centro
T\'ecnico Aeroespacial, 12228-900, S$\tilde{a}$o Jos\'e dos Campos,
Brazil}

\begin{abstract}
The lower bound masses of the ground-state relativistic three-boson 
system in 1+1, 2+1 and 3+1 space-time dimensions are obtained.
We have considered a reduction of the ladder Bethe-Salpeter
equation to the  light-front in a model with renormalized two-body
contact interaction. The lower bounds are deduced with the constraint
of reality of the two-boson subsystem mass.  It is verified that, in some
cases, the lower bound approaches the ground state binding energy. The
corresponding non-relativistic limits are also verified.
\end{abstract}
%%  \maketitle
%%  {\bf keywords:} 
%%  Bethe-Salpeter equation, Boson Systems, 
%%  Few body systems, dimensions  

\section{Introduction}                            

Recent applications of concepts of effective field theory 
to Quantum Chromodynamics, in which the effective
degrees of freedom are the constituent quarks, and the
baryons are viewed as bound system of three relativistic 
constituent quarks \cite{brodsky,karmanov}, make it worthwhile to 
study the properties of relativistic equations of three-particles
interacting through effective attractive forces. The schematic model,
represented by a contact interaction, gives one of the phenomenological
possibilities; even lacking confinement it can represent the
relativistic propagation of the system in a domain inside the 
confinement region. 
The relativistic equations can the reduced to the 
light-front~\cite{frede92,sales00}, allowing a definition of a 
light-front wave-function on the null-plane hypersurface ($x^+=t+z=0$). 
In the three-body case, the lower Fock-state component is given by a
three-particle wave-function. 
The radial wave-function of the baryon, for
light-quarks, is simplified by a totally symmetric component.
Disregarding spin effects, it can be represented by a three-boson 
wave-function.
To study the general properties of the three-boson Bethe-Salpeter 
equation, we allow the space-time dimensions to run from two to four 
($n=$2, 3 and 4), as in each case there are different physical aspects 
related to the dimensions that are relevant to address.

The relativistic light-front representation of the three-body
bound-state wave-function are well known to contain at least
two convenient and simple 
characteristics~\cite{karmanov,teren76,bakker79,frank79}:
\begin{itemize}
\item[a)] the center-of-mass coordinate (CM) can be easily separated; 
\item[b)] the wave-function is covariant under kinematical boosts. 
\end{itemize}
These features allow one to calculate some observables with the bound
state wave-function described only in terms of particle degrees of
freedom. Light-front wave-functions have been extensively used to
construct relativistic models for the light pseudoscalar 
mesons~\cite{dzim87,chung88,miller92,salme94,pache99}, 
vector mesons~\cite{keis94,pach97,jaus99} 
and for the
nucleon~\cite{brodsky,karmanov,konen90,chung91,salme95,pach95}, 
in terms of constituents quark bound systems.

The particular aspect that we are concerned in this work is the 
ground-state mass of the three-boson system in 2, 3 and 4 space-time
dimensions. The approach that we have 
considered consists of a relativistic Bethe-Salpeter equation 
reduced to the light-front, with renormalized two-body
contact interaction, parametrized by the two-boson binding energy. 
This model has been applied with some success to describe the 
charge distribution of the nucleon in terms of the constituent quarks
\cite{pach95}. Here, we derive a lower bound to the ground state binding
energy of a three-boson system, independent of the dimension, generalizing
the findings of reference \cite{frede92}, and compare with the numerical 
results from the solution of the Bethe-Salpeter equation. 
The approach does not have retardation effects, which simplifies 
the relativistic dynamical description of the bound state system.
The non-relativistic limit of the equation is also studied, since it
explains the results for the two-boson binding energies close to zero. 
 
The paper is organized as follows:
In section 2, we present the covariant ladder Bethe-Salpeter equation for
the Faddeev component of the three-boson vertex in $n-$dimensions
(Faddeev Bethe-Salpeter equation), using a zero-range two-body interaction
(a constant in momentum space). In a subsection, we also obtain explicitly
the corresponding two-body scattering amplitude, in the ladder
approximation, for space-time dimensions from two to four. As we have
indicated in this section and, subsequently, in section 3, the
generalization of the approach from zero-range to a finite range
interaction is not difficult.
In section 3, we perform the reduction of the equations to the
light-front, and obtain the corresponding non-relativistic limits. To make
the formalism more general, allowing the inclusion of finite range
effects, we include a single term separable interaction in the light-front
form of the equations for the vertex function.
We also discuss, in this section, the constraint in the phase-space of the
spectator particle momentum, due to the reality of the two-boson mass
subsystem. In section 4 we obtain the lower bounds to the masses of the
three-boson system (in the three cases we are considering). They are shown
to be related to the restricted phase-space of the spectator particle. 
Finally, in section 5, we present our numerical results and concluding 
remarks.

\section{Faddeev Bethe-Salpeter equation} 

The general features of the relativistic three-body equation projected in
the light-front have been discussed in ref.~\cite{bakker79}, where it was 
implicitly suggested a calculation with a specific interaction. Therefore,
it is instructive and illustrative to work out a simple dynamical model,
like the one we present here, which can be easily compared with the
corresponding covariant and non-relativistic approaches.
The model can also be considered in its present form
to define the dynamics in relativistic systems   in two, 
three and four space-time dimensions.

The Faddeev-like equation for the component of the three-boson
vertex is represented diagrammatically in figure 1, where 
the factor two in the rhs is related to the symmetrization of the 
total vertex. The vertex $v$, in the center-\-of-mass system, depends only
on the energy-momentum vector of the spectator particle, $q\equiv q^\mu$.  
The three-boson vertex is the sum of the three components in which
one of the bosons is the spectator. The two-boson scattering
amplitude, $\tau(M_2)$, enters in the kernel of the integral equation for
the vertex. From figure 1, we obtain
\begin{equation}
\! \! \! \! \! \! \! \! \! 
v(q^\mu) \! = \! 2 \tau(M_2)\int \! \frac{d^nk}{(2\pi)^n}
\left[\frac{\rm i}{k^2-M^2+i\varepsilon}\right]    
\left[\frac{\rm i}{(P_3-q-k)^2-M^2+i\varepsilon}\right]v(k^\mu), 
\label{eq:1}
\end{equation}
where $n$ is the dimension of the space-time,
$P^\mu_3$ is the three-boson energy-momentum vector in the
center-of-mass system ($P^0_3=M_{3B}$ and $P^{\mu\ne 0}_3=0$)
and the mass of the two-boson subsystem is given by
$M_2^2=(P_3-q)^2.$ \ 
In the present approach, the space-time dimension $n$ can be 
two, three and four, such that
the energy-momentum vector $q\equiv q^\mu$ is defined
as $(q^0,q^3)$ for $n=2$, $(q^0,q^3,q_\perp)$ for $n=3$, and
$(q^0,q^3,{\vec q}_\perp)$ for $n=4$.
$\vec{q_\perp}$ is a $(n-2)$ dimensional vector, that is 
zero for $n=2$.
%%%%%%%%%%%%%%  FIG.1 (EPS-FILE)   %%%%%%%%%%%%%%%%%%%%%
\begin{figure}
\begin{center}
\postscript{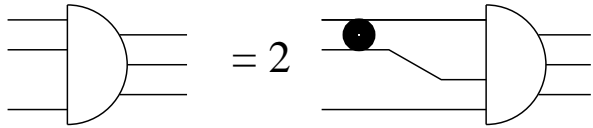}{0.5}
\caption{ 
Diagrammatic representation of the integral equation for the
Faddeev component of the vertex of three-boson bound-state.}
\end{center}
\label{fig.1}  
\end{figure}
%%%%%%%%%% EOFIG1

In the following subsection, the two-boson scattering amplitude
$\tau(M_2)$ that appears in eq.~(\ref{eq:1}) is explicitly obtained in the
light-front for two, three and four dimensions.

\subsection{Two-boson scattering amplitude}

The diagram in figure 2 shows the amplitude for the two-boson scattering
that, for a separable model interaction (as the present zero-range model)
can be solved analytically. The solution in the center-of-mass system is
given by
\begin{equation}  
\tau^{-1}(M_2)= 
{\rm i}\left\{\lambda^{-1} + {\rm i}B(M_2)\right\}, 
\label{eq:2}
\end{equation} 
where $\lambda$ is the coupling constant of the interaction
and $M_2$ the  mass of the two boson system. 
The factor {\rm i}, in eq.~(\ref{eq:2}), comes from the
$S-$matrix quantum field description of the two-body scattering
proccess, which is represented in Fig. 2 by the corresponding 
Feynman diagram.
$B(M_2)$ is the kernel of the scattering amplitude integral equation,
represented by the bubble diagram of figure 3, that is given by
\begin{equation} 
B(M_2)=-\int \frac{d^nk}{(2\pi)^n}\frac{1}
{\left\{\left(k^2-M^2+i\varepsilon\right)\left((P-k)^2-M^2+i\varepsilon\right)
\right\} },
\label{eq:3}
\end{equation}
where $M$ is the boson mass, $P^0=M_2$ and  $P^{\mu\ne 0}=0$. 

%%%%%%%%%%%%%%  FIG.2  %%%%%%%%%%%%%%%%%%%%%
\begin{figure}
\begin{center}
\postscript{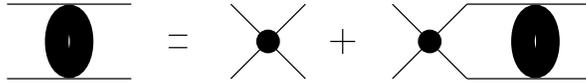}{0.5}
\end{center}
{\small
\caption{
Diagrammatic representation of the integral equation 
for the two boson scattering amplitude. 
}}
\label{fig.2}
\end{figure}
%%%%%%%%%% EOFIG2

%%%%%%%%%%%%%%  FIG.3 %%%%%%%%%%%%%%%%%%%%%
\begin{figure}
\begin{center}
\postscript{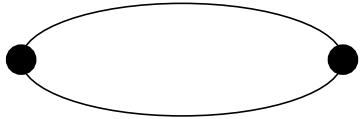}{0.3}
\end{center}
{\small \caption{
Bubble diagram corresponding to the
kernel of the integral equation~(\ref{eq:2}).
}}
\label{fig.3}
\end{figure}
%%%%%%%%%% EOFIG3

The momentum loop integral in eq. (\ref{eq:3}) is integrated in 
the null-plane
energy $k^-(=k^0-k^3)$,  $k^+(=k^0+k^3)$ and $\vec k_\perp$:
\begin{eqnarray}
B(M_2)= -\frac{1}{2(2\pi)^n} \int \frac{dk^-dk^+d^{(n-2)}k_\perp}{
k^+(P^+-k^+)} \left(k^--\frac{k^2_\perp(1- \delta_{n,2})+M^2-i\varepsilon}
{k^+}\right)^{-1}
\nonumber
\end{eqnarray}
\begin{eqnarray} 
\times \left(
P^-- k^- - \frac{(P-k)^2_\perp(1-\delta_{n,2}) +M^2-i\varepsilon}{P^+-k^+}
\right)^{-1},
\label{eq:4}
\end{eqnarray}
where the Kronecker symbol ($\delta_{i,j}=1, \delta_{i,{j\ne i}}=0$) was
introduced to take into account the two-dimensional case.

The finite integration over $k^-$ is performed with the use
of Cauchy theorem. The position of the poles in the $k^-$ complex-plane 
depends on the value of $k^+$. For $k^+ \ > \ P^+$ and  $ k^+ \ < \ 0$,
the poles are located in the same semi-plane, thus they don't contribute
to the integration.
Only if  $ 0 \ < \ k^+ \ < \ P^+$ the integration 
is non-vanishing. 
We use the fraction of momentum $x={k^+}/{P^+}$, within the
interval $ 0 \ < \ x \ < \ 1 \ . $ 
This condition yields the suppression of the antiparticle
propagation in the loop integral. 

The result for $B(M_2)$ is given by
\begin{eqnarray} 
{\rm i} B(M_2)= \frac{1}{2(2\pi)^{(n-1)}} 
\int \frac{dx}{x(1-x)}
\int \frac{d^{(n-2)}k_\perp}
{M^2_2 - \displaystyle\frac{k^2_\perp(1-\delta_{n,2})+M^2}{x(1-x)}} . 
\label{eq:5}
\end{eqnarray}

In the above equation, one may introduce a form factor to regularize the
transverse integration in four space-time dimensions. In order to keep the
Lorentz scalar property of the function $B(M_2)$, the form-factor can be a
function of the free mass of the two-boson intermediate state that, in
terms of the center-of-mass momentum, is written as 
\begin{equation}
M^{2}_{02}= 4(\vec{k}^2 + M^2) = \frac{k_\perp^2(1-\delta_{n,2})+M^2}
{x(1-x)}
\label{M02},
\end{equation} 
and
\begin{eqnarray} 
{\rm i} B(M_2)= \frac{1}{2(2\pi)^{(n-1)}} 
\int \frac{dx}{x(1-x)}
\int {d^{(n-2)}k_\perp} \frac{g^2(M_{02})}
{M^2_2 - M^2_{02}} . 
\label{B2}
\end{eqnarray}
A separable finite-range interaction between the two bosons
is characterized by the form-factor $g(M_{02})$, which is equal to one 
in the zero-range limit.

The value of $\lambda$ is chosen such that the two-boson system has
one bound-state. So, a pole occurs in the scattering amplitude at the
bound-state mass, where
\begin{equation}\lambda^{-1}= -{\rm i}B(M_{2B})
\label{lambda};\end{equation} 
and, from eq. (\ref{eq:2}), we obtain
\be 
\tau^{-1}(M_2)= B(M_{2B}) - B(M_2) . 
\label{eq:7} 
\ee
This result allows one to use $g\equiv 1$ in all the cases we are 
considering ($n=$2, 3 and 4), because the function $B(M_2)$ is 
finite for $n=2$ and $n=3$; and, for $n=4$, the divergence observed 
in eq.~(\ref{eq:5}) is canceled. 
In  four-dimensions, with $g=1$, the integral has a logarithmic
divergence that can be absorbed in the redefinition of  $\lambda$,
as in eq.(\ref{lambda}).  The physical information introduced in the
renormalization of the physical amplitude is the mass of the two bound
bosons, $ M_{2B}$.

The integration of $B(M_2)$ in two, three and four dimensions are
performed analytically. In two dimensions the two-boson
scattering amplitude for $M_2 \ < \ 2M$ is given by
\be
\tau^{-1}(M_2)= \frac{\rm i}{\pi} 
\left\{ 
\frac{{\rm arccot}\left[\beta_{2B}\right]} 
{M^{2}_{2B}\beta_{2B}}
- \frac{{\rm arccot}\left[\beta_{2}\right]}
{M^{2}_{2}\beta_{2}}
\right\}
,\label{eq:8}
\ee
where
\be 
\beta_{j}\equiv\sqrt{\left(\frac{2M}{M_{j}}\right)^2-1} \;\;\;\;
(j=2, 2B).
\label{eq:9}
\ee
In three dimensions ($n=3$), we have
\be 
\tau^{-1}(M_2)
= \frac{\rm i}{8\pi} \left\{ \frac{1}{M_{2B}} 
\ln \left(\frac{2M+M_{2B}}{2M-M_{2B}} \right)
-\frac{1}{M_{2}} \ln\left(\frac{2M+M_{2}}{2M-M_{2}} 
\right)
\right\}
\ ; 
\label{eq:10}
\ee
and, in four dimensions ($n=4$),
\be 
\tau^{-1}(M_2) = \frac{\rm i}{8\pi^2} 
\left[ \beta_{2B}\;
{\rm arccot} \left(\beta_{2B}\right)
- \beta_{2}\;{\rm arccot} \; 
\left(\beta_{2}\right)
\right]. 
\label{eq:tau4}
\ee
For the present purpose, the bound-state calculation, it is enough to know
$\tau(M_2)$ for $M_2 < 2M$. This concludes our discussion of the
renormalized zero-range two-boson interaction in the light-front. 
In case of a general two-boson separable interaction, with $g(M^2_{02})\ne
1$ in eq.~(\ref{B2}), one has to modify correspondingly the other
equations for the three-boson vertex function, with the inclusion of the
two-body form-factor, as shown in section 3. We also present 
in the next section the discussion of the phase-space constraints in
the light-front formalism of the equations.  

\section{Light-front reduction of the Bethe-Salpeter equation} 

The integral equation for $v$ is obtained after the
integration of equation (\ref{eq:1}) over $k^-$.
The integration over $k^-$ is performed with the use of Cauchy theorem,
assuming that the vertex is analytic in the lower half of the complex
$k^-$ plane. The pole, in the $k^-$ plane, comes from the 
on-mass-shell condition of the momentum of the spectator particle.
This allows direct access to the three-particle Fock-state component
of the wave-function. In the corresponding truncated Faddeev equation, 
only intermediate state propagations with three-particles are allowed.
Thus, in the vertex function, the spectator boson is on-mass-shell. 
The momentum variables in the integral equation,
$q^+$ and $q_\perp$, are the momentum in the null-plane for an
on-mass-shell particle (the transversal momentum is needed in three
and four space-time dimensions). 

We have started with a contact interaction for the bosons in 
eq. (\ref{eq:1}), which leads to a divergence in the transverse momentum
integration of the two-boson scattering amplitude in four space-time
dimensions. We also suggested the introduction of a separable interaction
with form-factor $g(M_{02})$, which by consistency should be present in
the kernel of the three-boson vertex equation due to the rearrangement of
the particles. As the form-factor depends on the free mass of the
intermediate interacting pair, it is natural to introduce the form-factors
as given below:

\begin{eqnarray}
v(y,\vec{p}_\perp) & = & \frac{{\rm i}\tau(M_2)}{(2\pi)^{n-1}}
\int \frac{dx d^{(n-2)}k_\perp}{x(1-y-x)}
\frac{g(M^A_{02}) g(M^B_{02})v(x,\vec{k}_\perp)}
{\left(M^2_{3B}-M^2_{03}\right)},
\label{vyp}
\end{eqnarray}
where 
\begin{eqnarray}
(M^A_{02})^2 &=& \frac{k^2_\perp+M^2}{x}+
\frac{(P_3-k-p)^2_\perp + M^2}{1-x-y} - (P_3-p)_\perp^2
\nonumber\\
(M^B_{02})^2 &=& \frac{p^2_\perp+M^2}{y}+
\frac{(P_3-k-p)^2_\perp + M^2}{1-x-y} - (P_3-k)_\perp^2 .
\end{eqnarray}
From now on, our discussions will be limited to the case of
contact interaction, with $g(M^2_{02})=1$.

Let us discuss the corresponding limits for the variables
$y\equiv{q^+}/{M_{3B}}$ and $q_\perp$. Following  ref.~\cite{frede92},
the mass of the two-boson subsystem is constrained to be real.
In two space-time dimensions, only the momentum fraction is enough to
describe the spectator boson, such that 
\be M^2_2\ = \
(M_{3B}-q^+)\left(M_{3B}-\frac{M^2}{q^+}\right) \ > \ 0 \ .  
\label{eq:11} \ee 
The above relation implies in the inequality 
\be 
1\ > \ y \ > \ \frac{M^2}{M^2_{3B}} \ . 
\label{eq:12} \ee 
In three and four space-time dimensions, the range of values of the
perpendicular momentum allowed by the reality of the mass of the two-boson
subsystem, comes from:  
\be M^2_2\ = \
(M_{3B}-q^+)\left(M_{3B}-\frac{q^2_\perp +M^2}{q^+}\right)-q^2_\perp \ > \
0 \ .
 \label{eq:13} \ee 
By solving this inequality for $q^2_\perp$, we obtain
\be q^2_\perp\ < \ (1-y)(M_{3B}^2y-M^2) .  
\label{eq:14} \ee 
The limits for $y$ are the same the dimensions $n=$2 and 3:  
\be 1 \ > \ y \ > \ \frac{M^2}{M^2_{3B}} \ , 
\label{eq:15}
\ee where the lower bound comes from $q^2_\perp \ > \ 0$. 

\subsection{Light-front Faddeev equation in $1+1$ space-time dimensions}

The equation for the Faddeev component of the vertex in $1+1$ space-time
dimensions is obtained using eq. (\ref{eq:8}) and
the limits given by eq. (\ref{eq:12}) together with the result of the
$k^-$ integration of eq. (\ref{eq:1}), as explicited in eq. (\ref{vyp})
with $g=1$:
\begin{eqnarray}
v(y) & = & - \left\{ 
\frac{2{\rm arccot}\left(\beta_{2B}\right)}
{M^{2}_{2B}\beta_{2B}}
- \frac{2{\rm arccot}\left(\beta_{2}\right)}
{M^{2}_{2}\beta_{2}}
\right\}^{-1} 
\nonumber \\ 
%%\end{eqnarray}
%%\begin{eqnarray}
%% \times 
& & \times \int^{1-y}_\frac{M^2}{M^2_{3B}}\frac{dx}{x(1-y-x)}
\frac{v(x)}{\left(M^2_{3B}-M^2_{03}\right)},
\label{eq:16}
\end{eqnarray}
where $M_2$ is given by eq.~(\ref{eq:11}), $\beta_2$ is defined in 
eq.(\ref{eq:9}) and the mass of the virtual three-boson state is:
\be 
M^2_{03}= M^2 \left(
\frac{1}{x}+\frac{1}{y} + \frac{1}{1-y-x}\right) \ . 
\label{eq:17}
\ee
The dependence of $v$ on $q^-$ is not specified considering that the
spectator boson is on mass-shell. $q^+$ describe the spectator boson
propagation.

Let us now study the non-relativistic limit  of eq.~(\ref{eq:16}) 
$(M\rightarrow \infty$). 
Considering that $M_2=2M-E_2$ and $M_{2B}=2M-E_{2B}$,  with
$E_{2B}$ the two-boson binding energy, 
in the limit $M\rightarrow \infty$ the two-boson amplitude
is given by:
\be \lim_{M\rightarrow \infty} \tau^{-1}=\frac{\rm i}{8 M^{3/2}}
\left[
\frac{1}{\sqrt{E_{2}}} - \frac{1}{\sqrt{E_{2B}}}
\right] \ . 
\label{eq:18}
\ee
The non-relativistic limit of the integrand of eq.~(\ref{eq:16}) 
involves the change in the variable $x$ to $k_z$:
\be 
\lim_{M\rightarrow \infty} \frac{dx}{dk_z}=
\frac{x}{M} \ , 
\label{eq:19}
\ee
and
\be 
\lim_{M\rightarrow \infty}x=\lim_{M\rightarrow \infty}y=
\frac13 \ . 
\label{eq:20}
\ee
As in eq.~(\ref{eq:16}) 
\be 
M_{03}=\sqrt{q^2+M^2}+\sqrt{k^2+M^2}
+\sqrt{( q+k)^2+M^2} \ , 
\label{eq:21}
\ee 
the non-relativistic limit is given by
\begin{eqnarray}
v_{NR}( q) & = &  \frac{2}{\pi \sqrt{M}} 
\left\{\frac{1}{\sqrt{E_{2}}}
-\frac{1}{\sqrt{E_{2B}}}\right\}^{-1} 
%%\\ \nonumber & & \times 
\int dk\frac{v_{NR}(k)}{-E_{3B}-\frac{q^2}{2M}
-\frac{k^2}{2M}-\frac{(q +k)^2}{2M}},  
\label{eq:22}
\end{eqnarray}                      
where $E_{3B}=3M-M_{3B}$ is the three-boson binding energy. 
This equation was obtained in ref.~\cite{dodd}. The nonrelativistic
binding energy is obtained analytically:
\be 
E_{3B}\ = \ 4E_{2B} \ . 
\label{eq:23}
\ee

\subsection{Light-front Faddeev equation in $2+1$ space-time dimensions} 

The corresponding equation for the Faddeev component of the vertex in 2+1
space-time dimensions is formulated from  eq.~(\ref{eq:10}) and
the limits given by eqs. (\ref{eq:14}) and  (\ref{eq:15})
together with the result of the $k^-$ integration of eq.
(\ref{eq:1}), as explicited by eq. (\ref{vyp}), which results:
\begin{eqnarray}
v(y,\vec q_\perp)&=&\frac{2}{\pi}
\left\{ \frac{1}{M_{2B}} \ln\left(\frac{2M+M_{2B}}{2M-M_{2B}}
 \right)
-\frac{1}{M_{2}} \ln\left(\frac{2M+M_{2}}{2M-M_{2}}
 \right)\right\}^{-1}  \nonumber \\
%%\end{eqnarray}
%%\begin{eqnarray}
&\times& \int^{1-y}_\frac{M^2}{M^2_{3B}}\frac{dx}{x(1-y-x)}
\int^{k_\perp^{max}}_{-k_\perp^{max}}dk_\perp
\frac{v(x,\vec k_\perp)}{M^2_{3B}-M^2_{03}}  \ , 
\label{eq:24}
\end{eqnarray}
where $M_2$ is given by eq.~(\ref{eq:13}), 
$$ k_\perp^{max}=\sqrt{(1-x)(M_{3B}^2x-M^2)}$$
and the mass of the virtual three-boson state is:
\be 
M^2_{03}=
\frac{k^2_\perp+M^2}{x}+
\frac{q^2_\perp+M^2}{y} + \frac{(q+k)^2_\perp +M^2}{1-y-x}  \  . 
\label{eq:25}
\ee
In this case, the non-relativistic limit $(M\rightarrow \infty)$
of the two-boson amplitude, eq. (\ref{eq:10}), is given by
\be \lim_{M\rightarrow \infty} \tau^{-1}=\frac{\rm i}{16 \pi M}
\left[\ln(E_{2}) -\ln(E_{2B})\right] \ , 
\label{eq:26}
\ee
where $M_{2B}=2M-E_{2B}$ and $M_2=2M-E_2$. 

The non-relativistic limit of the integrand of eq. (\ref{eq:24}) 
involves the change in the variable $x$ to $k_z$ in the
same steps as we did in 1+1 dimensions.
The mass of the virtual three-boson system is
substituted in eq. (\ref{eq:24}) by:
\be 
M_{03}=\sqrt{{\vec q}^2+M^2}+\sqrt{{\vec k}^2+M^2}
+\sqrt{(\vec q+ \vec k)^2+M^2} \ .
\label{eq:27}
\ee 
So, the non-relativistic limit is given by
\be 
v_{NR}( q) = \frac{2}{\pi M}\frac{1}{\left[\ln(E_{2B})
- \ln(E_2)\right]}
\int d^2k\frac{v_{NR}(k)}{-E_{3B}-\frac{{\vec q}^2}{2M}
-\frac{{\vec k}^2}{2M}-\frac{(\vec q + \vec k)^2}{2M}} \ , 
\label{eq:28}
\ee 
This is the non-relativistic equation  derived in 
ref. \cite{tjon}, with the nonrelativistic binding energy given by:
\be 
E_{3B}\ = \ 16.3 E_{2B} \ . 
\label{eq:29} 
\ee

\subsection{Light-front Faddeev equation in 3+1 space-time dimensions} 

The integral equation for the Faddeev component of the three boson
vertex, $v(y,\vec p_\perp )$, in the  center-of-mass system, is given by :
\begin{eqnarray}
v(y,\vec p_\perp)&\!=\!&\frac{1}{\pi} 
\frac{1}{\left[\beta_{2B} \; {\rm arccot} \left(\beta_{2B}\right)
- \beta_{2} \; {\rm arccot} \left(\beta_{2}\right)\right]} 
\nonumber \\
 & \times & \  \! \! \! \! \! \! 
\int^{1-y}_\frac{M^2}{M^2_{3B}}\frac{dx}{x(1-y-x)}
\int^{k_\perp^{max}}d^2k_\perp
\frac{v(x,\vec k_\perp)}{M^2_{3B}-M^2_{03}} \! \ . \!\!\!\!\! 
\label{eq:v4}
\end{eqnarray}
We assume that $ M_2 $, eq. (\ref{eq:14}),  has only real values
and, as a consequence, there is a maximum
of $ k_\perp^{max}=\sqrt{(1-x)(M_{3B}^2x-M^2)}$ .

Let us now study the non-relativistic limit of eq. (\ref{eq:v4}) 
$(M\rightarrow \infty$). In this limit, from eq. (\ref{eq:tau4}) we have:
\be \lim_{M\rightarrow \infty} \tau^{-1} =
\frac{\rm i}{16\pi \sqrt{M}} 
\left[\sqrt{E_{2B}}-\sqrt{E_2}\right] \ ,
\label{eq:tau4nr}
\ee
where $M_{2B}=2M-E_{2B}$ and $M_2=2M-E_2$. 

The non-relativistic limit of the integrand of eq. (\ref{eq:v4}) 
involves the change in the variable $x$ to $k_z$ in the same steps as we
did in 2+1 dimensions. From eqs. (\ref{eq:27}) and (\ref{eq:v4}), 
\begin{equation}
v_{NR}(\vec{q})=\frac{\sqrt{M}}{\pi^2(\sqrt{E_{2B}} - \sqrt{E_2})}\int
\frac{d^{3}{k}}{M^2}\frac{v_{NR}(\vec{k})}
{\left[-E_{3B}-\frac{\vec{q}^2}{2M}-
\frac{\vec{k}^2}{2M}-\frac{(\vec{q}+\vec{k})^2}{2M}\right]}
\label{eq:espnr} \ .
\end{equation}
In eq. (\ref{eq:v4}), the Faddeev component of the ground state vertex 
is rotationally symmetric in the perpendicular plane,
while in eq. (\ref{eq:espnr}) it is spherically symmetric. 

The Efimov effect \cite{efimov70} appears in  eq.~(\ref{eq:v4}).
In the Efimov limit, that is reached when the two-body binding energy is
zero (infinite scattering length), the nonrelativistic Faddeev
equation~(\ref{eq:espnr}) in the $s-$wave presents an infinite 
number of three-body bound-states close to zero. 
The same limit can be identified with the Thomas collapse~\footnote{In
1935, Thomas~\cite{thomas35} has shown that the triton binding energy will
collapse in the zero-range limit of the two-body interaction.}, through a
reescaling of the equation (\ref{eq:espnr}).
In order to solve the eq.~(\ref{eq:espnr}), we need to use a cutoff
$\Lambda$ in the momentum integration to avoid the collapse of the
three-boson system. In units such that $\Lambda=1$ and $M=1$, this 
equation can be written as 
\begin{equation}
v_{NR}(\vec{z})=\frac{1}{\pi^2(\sqrt{\varepsilon_{2B}} - 
\sqrt{\varepsilon_2})}\int
{d^{3}{z^\prime}}\frac{v_{NR}(\vec{z}^\prime)\theta(1-|\vec{z}^\prime|)}
{\left[-\varepsilon_{3B}-\frac{z^2}{2}-
\frac{z^{\prime 2}}{2}-\frac{(\vec{z}+\vec{z}^\prime)^2}{2}\right]}
\label{eq:espnr2} \ ,
\end{equation}
where 
$\varepsilon_{N} = E_{N}/\Lambda^2$ ($N=2$, $2B$ or $3B$).
The collapse of the three-boson system occurs for $\Lambda\to\infty$, when
$E_2$ is fixed. The Efimov effect is manifested for $E_{2B}\to 0$, when
$\Lambda$ is fixed. So, in the nonrelativistic system both effects are
related by a scale transformation and correspond to the same limit
$\varepsilon\to 0$ (see also ref.\cite{frede88}).

The Efimov effect holds in the relativistic approach considered
in this paper, but the short-range behavior is strongly modified:
the Thomas collapse~\cite{thomas35} of the system is avoided, 
due to the lower bound of the ground state mass in eq.~(\ref{eq:v4}),  
as it will be discussed in the next section.

\section{Lower bounds to the three-boson binding energy} 

Returning to the relativistic equations in $1+1$, $2+1$ and $3+1$ 
dimensions, eqs. (\ref{eq:16}), (\ref{eq:24}) and
(\ref{eq:v4}), respectively, the lower bound for the mass of the
three-boson system is obtained from the limits on the $x$ integration and
the condition $y>{M^2}/{M^2_{3B}}$, which implies:
\be 
\frac{M^2}{M_{3B}^2} < 1 -\frac{M^2}{M_{3B}^2},
\ee
giving the bound
\be
M_{3B}> \sqrt{2}M \ . 
\label{eq:30} 
\ee
The same inequality has been already discussed in $3+1$ dimensions 
in reference \cite{frede92}.
The following inequality holds $M+M_{2B}>M_{3B}> \sqrt{2}M $; and, by 
using $M_{2B}=2M-E_{2B}$ and $M_{3B}=3M-E_{3B}$, we obtain
$E_{2B}<E_{3B}<(3-\sqrt{2})M$. 
Defining the binding energy in respect to the elastic scattering threshold
of one boson and the two-boson bound state,
$B_{3B}=M+M_{2B}-M_{3B}$, the lower bound for the ratio between $B_{3B}$
and the two-boson binding energy is given by
\begin{equation}
0<\frac{B_{3B}}{E_{2B}}<(3-\sqrt{2})\frac{M}{E_{2B}}-1 \ .
\label{eq:lb}
\end{equation}
The ratio between the three-boson binding energy and the two-boson binding
energy is taken to stress the universal scaling  that occurs
in the non-relativistic limit, in 1+1 and 2+1 space-time dimensions.
However, in the relativistic domain, the boson mass is also an energy
scale which is evidenced by the lower bound in eq. (\ref{eq:lb}). The
numerical results confirm the boson mass as a physical scale. 

In 3+1 space-time dimensions, the non-relativistic limit of the three-boson
system, has the collapse of the ground state if no physical
cutoff to high momentum intermediate state propagation
in eq. (\ref{eq:espnr}) exists. The two-boson binding energy in the strict
non-relativistic limit is not the only physical scale and the collapse
represents the emergence of a three-body scale. However, in the 
relativistic eq. (\ref{eq:v4}), the boson mass scale matters
and the collapse is avoided. As $E_{2B}$ goes to zero the
Efimov effect is still present in the relativistic eq. (\ref{eq:v4}),
as the boson mass becomes irrelevant as a physical momentum scale in 
the non-relativistic limit \cite{frede92}.
In nonrelativistic quantum mechanics the Efimov and Thomas effects
are equivalent phenomena as one can verify by rescaling the two-body 
scattering length and interaction range \cite{frede88}, however 
the relativistic propagation of the bosons breaks the equivalence 
as the boson mass is a momentum scale that has no effect in the
nonrelativistic limit, but avoids the collapse in the relativistic
situation~\cite{frede92,perez}.

\section{Numerical results and conclusion}

In figure 4, for $M=1$, the numerical solution of the light-front 
Faddeev equations for the ratio between the
binding energies of  three-boson to the two-boson
systems is shown as a \linebreak function of the two-boson binding energy
for $E_{2B}<(3-\sqrt{2})$ and compared
to the lower bound \linebreak $(3-\sqrt{2})/{E_{2B}}-1 $ . 
In (1+1) dimensions, the non-relativistic limit of 
eq. (\ref{eq:16}) for $E_{2B} << 1$, results in the nonrelativistic
equation for the zero-range potential, eq. (\ref{eq:22}), which has
a solution for $B_{3B}/E_{2B}=3$ \cite{dodd}, as is observed in figure 4. 
In (2+1) dimensions, the non-relativistic limit of 
eq. (\ref{eq:24})  is the nonrelativistic equation
for the zero-range potential, eq. (\ref{eq:28}), which has a solution for
$B_{3B}/E_{2B}=15.3$ \cite{tjon}, that  we can see  in figure 4 as $E_{2B}$ 
approaches zero.
In (3+1) dimensions, the light-front three-boson eq. (\ref{eq:v4}), even
being formally similar to the nonrelativistic eq. (\ref{eq:espnr}),
still carries the boson mass scale, due to the momentum constraints in 
eq. (\ref{eq:v4}), and consequently $B_{3B}$ goes to a finite value as
$E_{2B}$ is close to zero.
It does not present the Thomas effect. The lower bound of $B_{3B}$ is 
approached by  our calculation in (1+1) and (2+1) dimensions for $E_{2B}$
above $0.4$, while it
overestimate the binding in (3+1) dimensions by a factor of about 3.

%%%%%%%%%%%%%%  FIG.4 (EPS-FILE) %%%%%%%%%%%%%%%%%%%%%
\begin{figure}
\label{fig4}  
\postscript{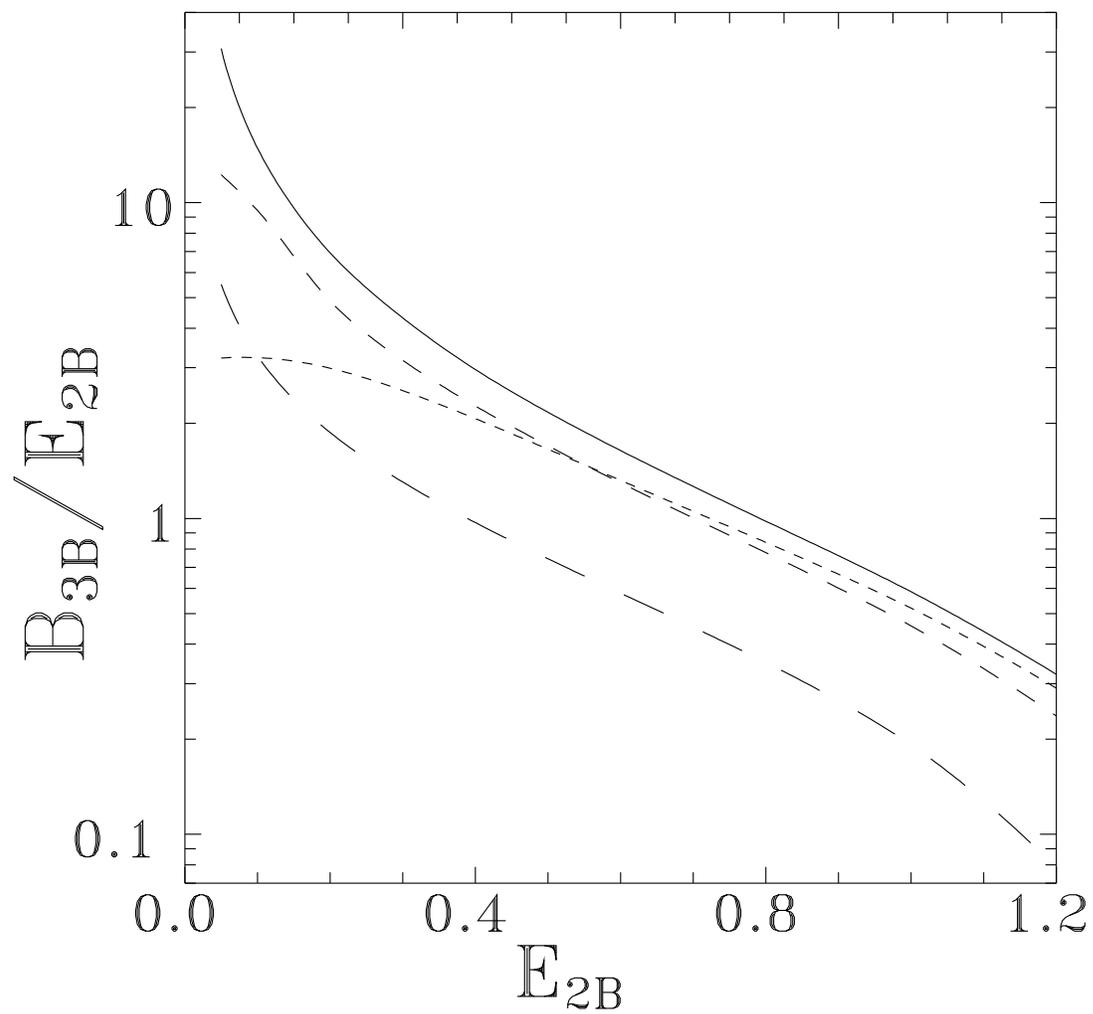}{1.0} 
\caption{
The three-boson binding energy, $B_{3B}=M+M_{2B} - M_{3B}$, is represented 
as a function of the two-boson mass in units of the single-boson mass,
for one (short-dashed-line), two (dashed-line) and three
(long-dashed-line) space-time dimensions.
The solid curve is the lower bound.}
\end{figure}
%%%%%%%%%%%%%%%%   EOFIG4

In conclusion, we show how to reduce the ladder Bethe-Salpeter
equation for a three-boson system to the null-plane. The 
light-front dynamics of the three-boson bound state 
was performed in $1+1$, $2+1$ and $3+1$ space-time dimensions. 
The two-boson bound-state mass is used to parametrize the mass of the
virtual two-body subsystem, in the renormalized two-body
amplitude of the relativistic three-body Faddeev equation.
With the mass of the virtual two-body subsystem constrained to real values, 
the momentum phase-space of the spectator particle is delimited, which
implies on a lower bound to the three-body ground state energy. 
We compared the ground state energy obtained from the numerical solution
of the light-front Faddeev equations and found that the lower bound
is approached by the system ground state energy in (1+1) and 
(2+1) space-time dimensions in the relativistic region of the two-boson
bound state energy, where it attains values around the boson mass.
Since the lower bound derivation involves only phase space considerations, 
it seems possible to extend the present work to three-fermions systems 
in the light-front interacting with contact  interactions. 

\section*{Acknowledgements}
This work was partially supported  by the Conselho Nacional de
Desenvolvimento Cientifico e Tecnol\'ogico - CNPq and Funda\c c\~ao de
Amparo \`a Pesquisa do Estado de S\~ao Paulo - FAPESP.


\section*{References}
\begin{thebibliography}{99}  

\bibitem{brodsky} S.J. Brodsky, H.-C. Pauli, S.S. Pinsky 
1998 {\it Phys. Rep.} {\bf 301} 299.

\bibitem{karmanov} J.\ Carbonell, B.\ Desplanques, V.\ Karmanov and
  J.-F.\ Mathiot 1998 {\it Phys.\ Reports} 
  {\bf 300} 215   and references therein.

\bibitem{frede92} T.Frederico 1992 {\it Phys. Lett.} {\bf B 282} 409.

\bibitem{sales00} J.H.O. Sales, T.Frederico, B.V. Carlson and P.U.
Sauer 2000 {\it Phys. Rev.} {\bf C 61} 044003.

\bibitem{teren76}M. V. Terent'ev 
1976 {\it Sov. J. Nucl. Phys.} {\bf 24} 106;
L. A. Kondratyuk and M.V.Terent'ev 1980 
 Sov. J. Nucl. Phys. {\bf 31} 561.

\bibitem{bakker79} B.L.G. Bakker, L.A.Kondratyuk and 
M.V.Terent'ev 1979 {\it Nucl. Phys.} {\bf B 158} 497.

\bibitem{frank79} L.L.Frankfurt and M.I.Strikman 1979 Nucl. Phys. 
{\bf B 148} 107; ibid., 1981 {\it Phys. Rep.} {\bf 76} 215.

\bibitem{dzim87} Z. Dziembowski and L. Mankiewicz 1987  
{\it Phys. Rev. Lett.} {\bf 58} 2175; 
ibid. 1988 {\it Phys. Rev.} {\bf D 37} 778.
 
\bibitem{chung88} P. L. Chung , F. Coester and 
W. N. Polizou 1988 {\it Phys. Lett.} {\bf B 205} 545; 
1990 C.R. Ji and S. Cotanch, {\it Phys. Rev.} {\bf D 21} 2319;
I. G. Aznauryan and K. A. Oganesyan 1988 
{\it Sov. J. Nucl. Phys.} {\bf 47} 1097;
ibid. 1990 {\it Phys. Lett.} {\bf B 249} 309.

\bibitem{miller92}T. Frederico and G. Miller 
1992 {\it Phys. Rev.} {\bf D 45} 4207, 
ibid. 1994 {\bf D 50} 210.

\bibitem{salme94}F. Cardarelli, I.L. Grach, 
I.M. Narodetski, E. Pace, G. Salme and 
S. Simula 1994 {\it Phys. Lett.} {\bf B 332} 1.

\bibitem{pache99}J.P.B.C. de Melo, H.W.L. Naus and 
T. Frederico 1999 {\it Phys. Rev.} {\bf C 59} 2278.

\bibitem{keis94}B. D. Keister 1994 
{\it Phys. Rev.{\bf D 49}} 1500.

\bibitem{pach97} J.P.B.C. de Melo and 
T.Frederico 1997 {\it Phys.Rev.} {\bf C 55} 2043.

\bibitem{jaus99}W. Jaus 1999 {\it Phys. Rev.} {\bf D 60} 054026.

\bibitem{konen90}W.Konen and H.J. Weber 
1990 {\it Phys. Rev.} {\bf D 41} 2201.

\bibitem{chung91} P.L.Chung and F. Coester 1991 
{\it Phys. Rev.} {\bf D 44} 229.

\bibitem{salme95}F. Cardarelli, E. Pace, G. Salme and S. Simula 
1995 {\it Phys. Lett.} {\bf B 357} 267; 

1995 {\it Few Body Syst. Suppl.} {\bf 8 } 345; 
F. Cardarelli and S. Simula, nucl-th/9906095, 
1999 {\it Phys. Lett.}  {\bf B 467} 1.

\bibitem{pach95} W.R.B. de Ara\'ujo, J.P.B.C. de Melo and 
T.Frederico 1995 {\it Phys.Rev.} {\bf C 52} 2733.

\bibitem{dodd} L.R. Dodd 1970 {\it J. Math.} {\bf 11} 207.

\bibitem{tjon} L.W. Bruch  and  
J. A. Tjon 1979 {\it Phys. Rev.} {\bf A 19} 425.

\bibitem{efimov70}V. Efimov 1970 {\it Phys. Lett.} {\bf B 33} 563; 
see also R. D. Amado and 
J. V. Noble 1972 {\it Phys. Rev.} {\bf D 5} 1992.

\bibitem{thomas35} L.H. Thomas 1935 {\it Phys. Rev.} {\bf 47} 903.

\bibitem{frede88}S. K. Adhikari, A. Delfino, T. Frederico, I. D. Goldman
  and  L. Tomio 1988 {\it Phys. Rev.} {\bf A 37} 3666.

\bibitem{perez}F.A.B. Coutinho, J.F. Perez, 
W.F. Coutinho 1999 {\it Ann. Phys.} {\bf 277} 94.
\end{thebibliography}
\end{document}